\documentclass{PoS}
\pdfoutput=1
\usepackage{amsmath}
\usepackage{epsfig}

\title{On the impact of off-shell contributions on the top quark mass extraction in $t\bar{t}j$ events}

\ShortTitle{On the impact of off-shell contributions on the top quark mass extraction}

\author{\speaker{Manfred Kraus}\thanks{Preprint number: HU-EP-17/26}\\
Humboldt-Universit\"at zu Berlin\\ Institut f\"ur Physik\\ Newtonstra\ss e 15\\
D-12489 Berlin, Germany\\
E-mail: \email{manfred.kraus@physik.hu-berlin.de}
}

\abstract{In this contribution we present a systematic comparison
of the full off-shell calculation for $pp\to t\bar{t}j$ in the dileptonic
decay channel with a description based on
the narrow width approximation. The aim of this study is to estimate the size
of off-shell and non-resonant contributions 
to the shape of differential distributions, that will be utilized in the
top quark mass extraction using the template fit method.
}

\FullConference{13th International Symposium on Radiative Corrections\\
		24-29 September, 2017\\
		St. Gilgen, Austria}

\begin{document}

\section{Introduction}
With the advent of the LHC Run 2 at a center-of-mass energy of $\sqrt{s}=13$ TeV
top quark physics entered the precision era. The main goals of the top quark
physics program are the precise determination of top quark properties, such as
its mass, its coupling to the Higgs boson as well as gauge bosons. Other key
measurements include differential distributions, fiducial cross sections and
the measurement of spin correlations of the top quark decay products.

For now the most precise calculations for stable top quarks are the total cross
section at NNLO+NNLL~\cite{Czakon:2013goa} and differential distributions at
NNLO QCD + NLO EW~\cite{Czakon:2017wor}. However, the top quark is a
highly unstable particle and decays before it hadronizes and can thus be studied
via its decay products. Including the top quark decay in the calculation allows
the precise study of fiducial phase space regions and offers a closer modelling
of the experimentally accessable final states. To this end, the decay has been
incorporated in the narrow width approximation at NLO in
Refs.~\cite{Melnikov:2009dn,Campbell:2012uf} and at approximate NNLO in
Ref.~\cite{Gao:2017goi}. Contrary to this approach, in Refs.
\cite{Denner:2010jp,Bevilacqua:2010qb,Denner:2012yc,Heinrich:2013qaa,
Frederix:2013gra,Cascioli:2013wga}
the on-shell treatment of top quarks has been abandoned and the complete NLO QCD
corrections for the process $pp \to e^+\nu_e\mu^-\bar{\nu}_\mu b\bar{b} + X$
were calculated and are now also consistently matched with parton showers
\cite{Jezo:2016ujg}. Recently also the NLO EW corrections have become
available~\cite{Denner:2016jyo} as well as the NLO QCD corrections for the
semi-leptonic decay channel~\cite{Denner:2017kzu}.

\begin{figure}
\begin{center}
	\includegraphics[width=0.6\textwidth]{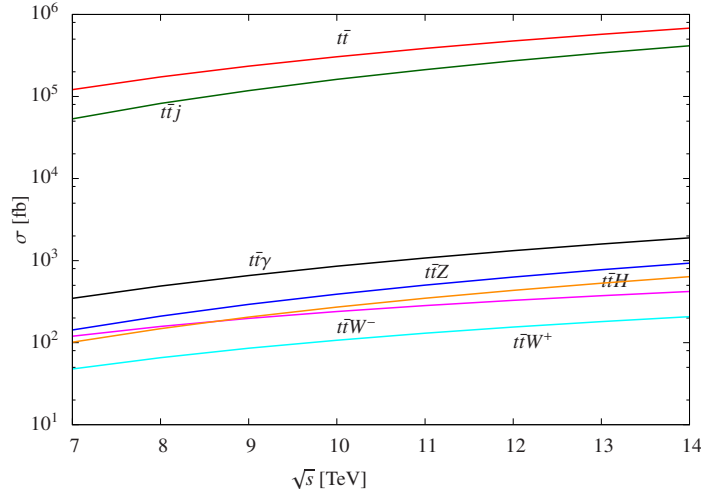}
	\caption{Production cross section of $t\bar{t}$ and associated $t\bar{t}$
	production as a function of the center-of-mass energy $\sqrt{s}$. }
	\label{fig:propaganda}
\end{center}
\end{figure}

However, at the energies that the LHC is operating, top quarks are abundantly
produced in association with either additional jets or electroweak bosons as
illustrated in Fig~\ref{fig:propaganda}.
As one can see, a considerable amount of top quark pair events are actually
accompanied by an additional hard jet. Therefore, we want to focus on the
$pp \to t\bar{t}j$ process which has already been studied in great detail in recent
years~\cite{Dittmaier:2007wz,Dittmaier:2008uj,Melnikov:2010iu,Melnikov:2011qx,
Kardos:2011qa,Alioli:2011as,Czakon:2015cla,Bevilacqua:2015qha,
Bevilacqua:2016jfk}.
The $pp \to t\bar{t}j$ process is particulary interesting because
it can be used to extract the top quark mass parameter.
In Refs.~\cite{Alioli:2013mxa,Fuster:2017rev} an observable has been designed
for this process that has larger $m_t$ sensistivity than the inclusive top quark
pair production process. This observable has been also already sucessfully
used by the experimental collaborations~\cite{Aad:2015waa,CMS:2016khu} to
extract the top quark mass.
Here, we present parts of our recent study~\cite{Bevilacqua:2017ipv} that
extends the aforementioned $t\bar{t}j$ studies where we address the impact
of off-shell effects on the extraction of the top quark mass parameter.

\section{Outline of the calculation}
We want to address the size of off-shell effects on the top quark mass
extraction from differential distributions. To this end, we compare the full
off-shell calculation for the $pp \to e^+\nu_e\mu^-\bar{\nu}_\mu b\bar{b}j$
final state with different approximations of the calculation. The NLO QCD
calculation, as described in detail in Refs.~\cite{Bevilacqua:2015qha,
Bevilacqua:2016jfk}, is used and takes into account all double, single and
non-resonant contributions as well as their interference contributions at
$\mathcal{O}(\alpha_s^4\alpha^4)$. This calculation has been performed within
the \textsc{HELAC-NLO} framework~\cite{Bevilacqua:2011xh}, which consists of
the two building blocks \textsc{HELAC-DIPOLES}~\cite{Czakon:2009ss,
Bevilacqua:2013iha} and \textsc{HELAC-1LOOP}~\cite{vanHameren:2009dr}.

The full off-shell calculation is compared to the narrow-width-approximation
(NWA) as presented in Ref.~\cite{Melnikov:2011qx}, where the matrix element can
be factorized into the following on-shell top quark contributions
\begin{equation}
\begin{split}
\lim_{\Gamma_t/m_t \to 0}~\left|M^{WWb\bar{b}j}\right|^2 &=  \left| M^{t\bar{t}j}\right|^2
   \otimes Br(t\to Wb) \otimes Br(\bar{t}\to W\bar{b})  \\
&+ \left| M^{t\bar{t}} \right|^2 \otimes Br(t\to Wbj) \otimes Br(\bar{t}\to W\bar{b}) \\
&+ \left| M^{t\bar{t}} \right|^2 \otimes Br(t\to Wb) \otimes Br(\bar{t}\to W\bar{b}j)
  + \mathcal{O}\left(\frac{\Gamma_t}{m_t}\right)\;,
\label{eqn:NWA}
\end{split}
\end{equation}
where we ignored for brevity the leptonic decays of the $W$ bosons.
We also compare the full off-shell calculation with an approximation dubbed
\textit{NWA$_{prod}$} that employs NLO QCD corrections only to the production
part while taking only leading order (LO) top quark decays into
account and represents the calculation presented in Ref.~\cite{Melnikov:2010iu}.

The calculations are performed for the LHC at a center-of-mass energy of
$\sqrt{s}=13$ TeV, while the detailed list of standard model input parameters,
phase space cuts, etc. can be found in Ref.~\cite{Bevilacqua:2017ipv}.
The CT14~\cite{Dulat:2015mca}, MMHT14~\cite{Harland-Lang:2014zoa} and
the NNPDF 3.0~\cite{Ball:2014uwa} PDF sets have been employed in the calculation,
while a common renormalization and factorization scale $\mu_R=\mu_F=\mu_0$ has
been chosen. We consider three different scale choices, a fixed scale
$\mu_0 = m_t$ and two dynamical ones, namely $\mu_0 = E_T/2$ and
$\mu_0 = H_T/2$, with
\begin{equation}
 E_T = \sqrt{m_t^2 + p^2_T(t)} + \sqrt{m_t^2 + p_T^2(\bar{t})}\;,
\end{equation}
where the top quark momenta are reconstructed from the final state momenta, i.e.
the top quark momentum is given by $p(t) = p(e^+) + p(\nu_e) + p(j_b)$. On the
other end, $H_T$ is independent of the underlying process and defined as
\begin{equation}
  H_T = p_T(e^+) + p_T(\mu^-) + p_T(j_b) + p_T(j_{\bar{b}}) + p_T(j_1)
  + p_T^{miss}\;.
\end{equation}
As already mentioned in the introduction, we focus in this contribution on two
particular observables, namely $\mathcal{R}(m_t^{pole},\rho_s)$ and $M_{be^+}$.
Before, we discuss the top quark mass extraction from these
distributions in detail, we will investigate them closer and discuss shape differences
between the different approaches.

Our first observable under investigation, $\mathcal{R}(m_t^{pole},\rho_s)$, is
defined by
\begin{equation}
  \mathcal{R}(m_t^{pole},\rho_s) = \frac{1}{\sigma_{t\bar{t}j}}
  \frac{d\sigma_{t\bar{t}j}}{d\rho_s}(m_t^{pole},\rho_s)\;, \qquad \text{with}
  \qquad   \rho_s = \frac{2m_0}{M_{t\bar{t}j}}\;,
\end{equation}
where $m_0=170$ GeV is an arbitrary scale and $M_{t\bar{t}j}$ the invariant
mass of the $t\bar{t}$ system and the leading hard jet.
\begin{center}
\begin{figure}
  \includegraphics[width=0.9\textwidth]{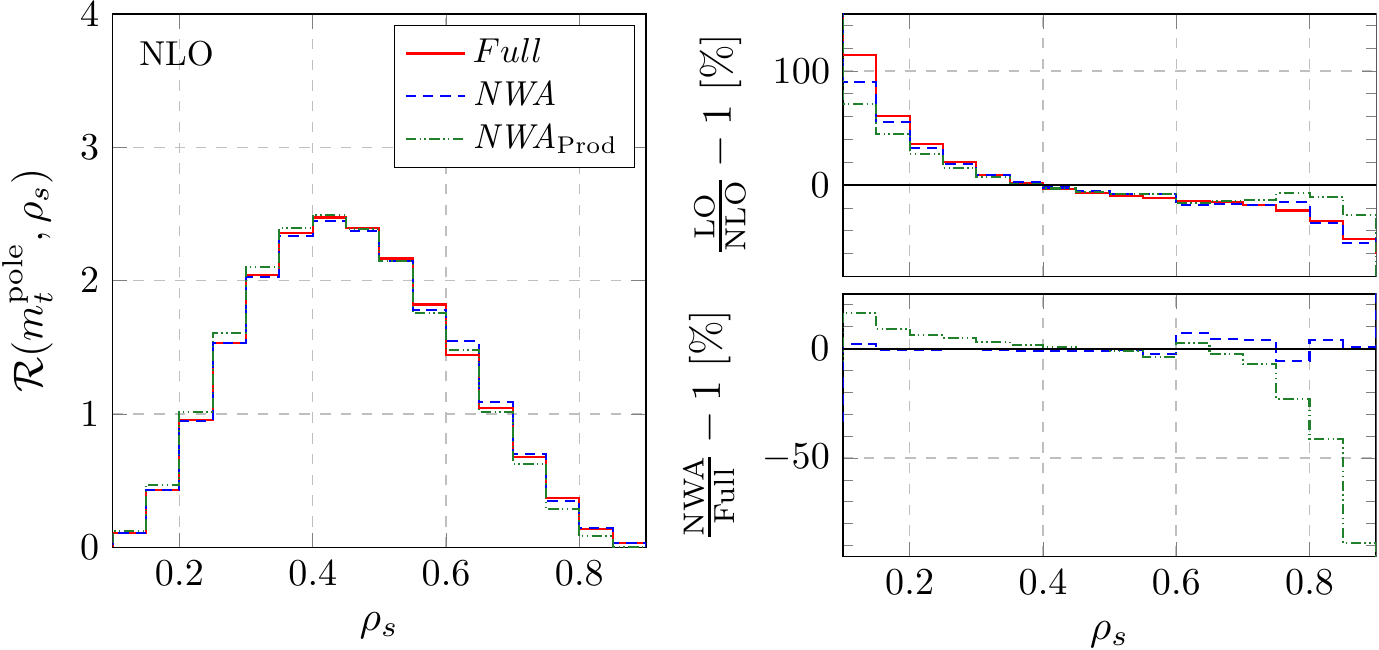}
  \caption{The normalized differential distribution for
  $\mathcal{R}(m_t^{pole},\rho_s)$ for the LHC at $\sqrt{s}=13$ TeV with
	$\mu_0 = m_t = 173.2$ GeV.}
  \label{fig:rhocompare}
\end{figure}
\end{center}
In Fig.~\ref{fig:rhocompare} the $\mathcal{R}(m_t^{pole},\rho_s)$ distribution
for the three calculations mentioned above together with their NLO K-factors and
the relative deviation from the full off-shell calculation is shown.
Let us first note, that in all three cases the NLO K-factor is not flat and the
shape is altered up to $50\%$ for \textit{Full} and \textit{NWA} in the
threshold region ($\rho_s \approx 1$).
On the other hand, the \textit{NWA} result approximates the \textit{full}
off-shell calculation reasonably well over the whole spectrum of the distribution
and differences of at most $15\%$ are visible in the region of
$\rho_s > 0.6$. Contrary, \textit{NWA$_{prod}$} yields substantial
differences of up to $85\%$ in the threshold region. These differences will
have a sizeable impact on the top quark mass parameter as we will show later.

\begin{center}
\begin{figure}
  \includegraphics[width=0.9\textwidth]{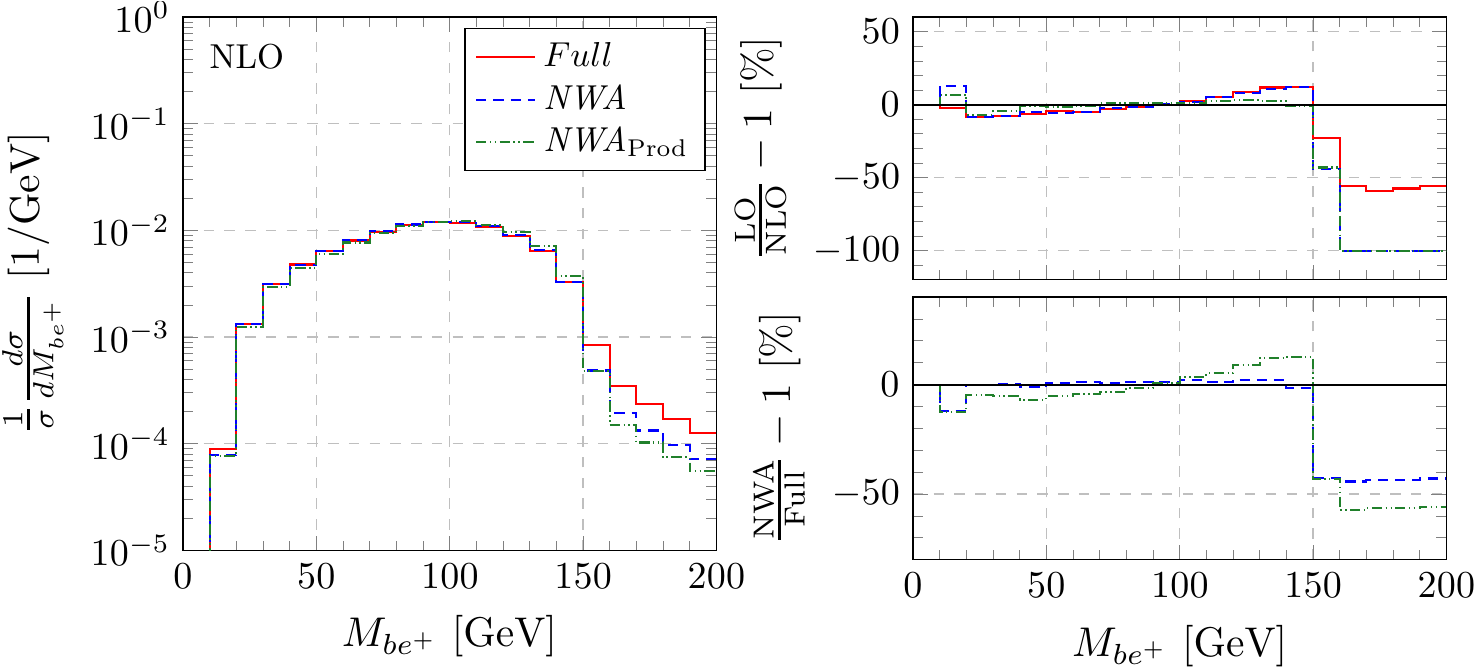}
  \caption{The normalized differential distribution for
  $M_{be^+}$ for the LHC at $\sqrt{s}=13$ TeV with $\mu_0 = m_t = 173.2$ GeV.}
  \label{fig:mblcompare}
\end{figure}
\end{center}

The second observable under investigation, a rather standard one, is the
normalized differential distribution of $M_{be^+}$, which is defined as
\begin{equation}
 \frac{1}{\sigma_{t\bar{t}j}} \frac{d\sigma_{t\bar{t}j}}{dM_{be^+}}\;, \qquad
 \text{where}\qquad M_{be^+} = \min\{M_{b_1e^+}, M_{b_2e^+} \}\;.
\end{equation}
This observable has a kinematical endpoint at $M^{max}_{be^+} =
\sqrt{m_t^2-m_W^2} \approx 154$ GeV, if top quarks and W bosons are considered
on-shell. Therefore, only additional radiation as well as off-shell effects
can smear this boundary. The observable is presented in Fig.~\ref{fig:mblcompare}
where we see that neglecting the QCD corrections to the top quark decay
amounts to shape differences between \textit{NWA$_{prod}$} and \textit{Full} of the order of $15\%$ at the kinematical endpoint. On the other hand, \textit{NWA}
accounts for all dominant contributions below the kinematical endpoint. Thus,
below the endpoint \textit{NWA} and \textit{Full} are essentially identitical.
However, above the endpoint we see deviations from the full off-shell
calculation of the order of $50\%$. Nonetheless, due to the severe drop of the
cross section above the kinematical endpoint these large off-shell effects do
not impact the top quark mass extraction substantially.
%
\section{Top quark mass extraction}
\begin{figure}
  \includegraphics[width=0.52\textwidth]{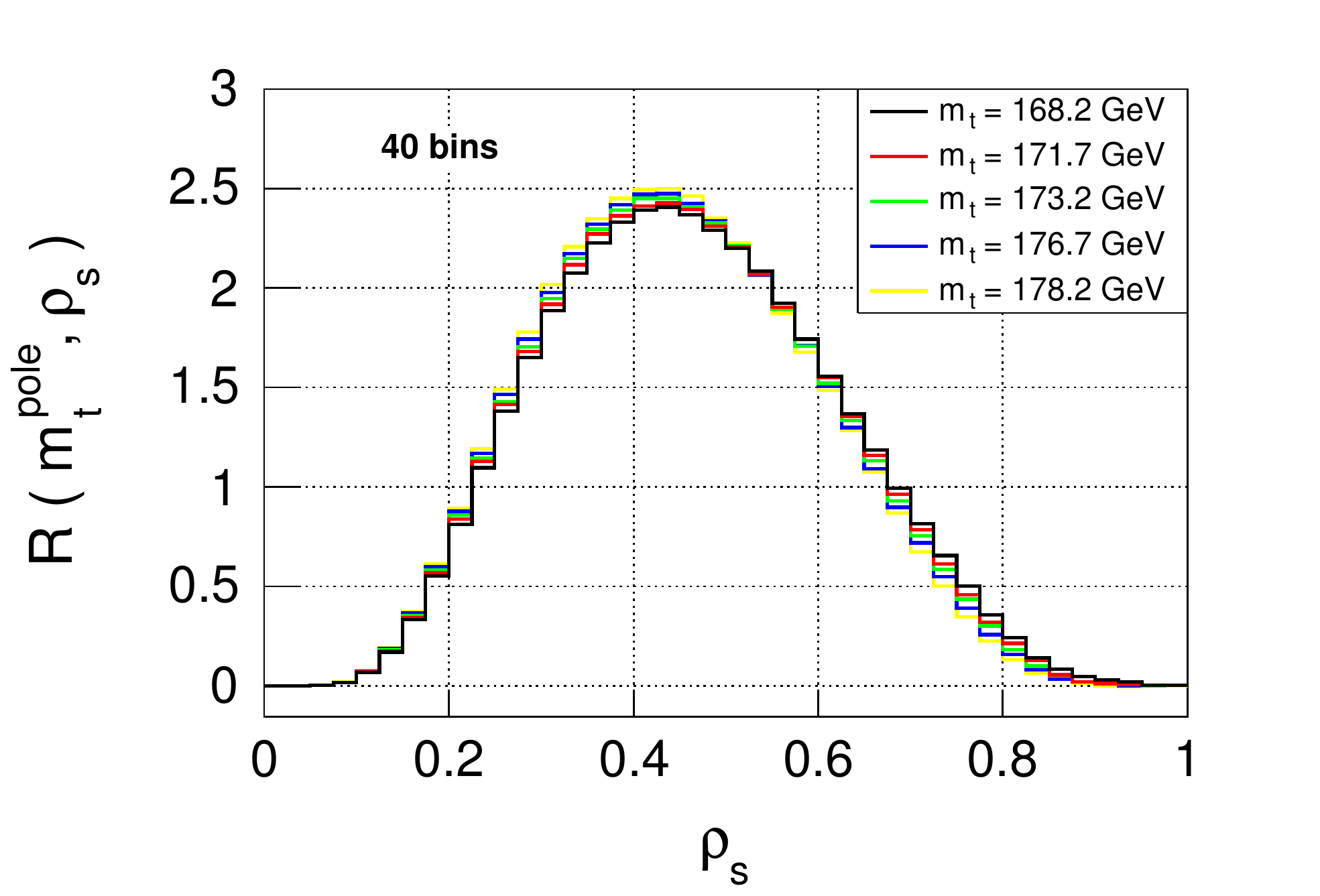}
  \includegraphics[width=0.48\textwidth]{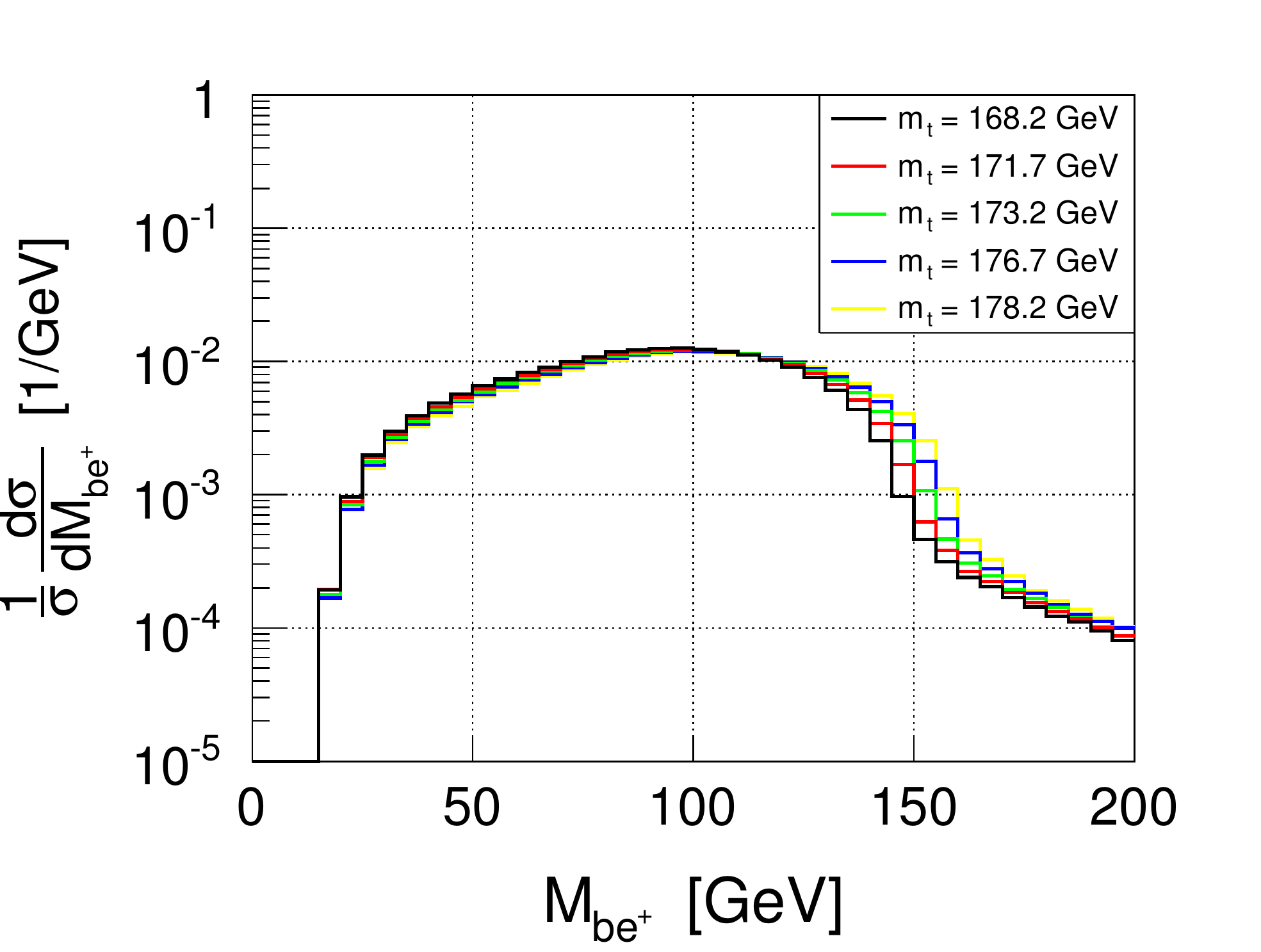}
  \caption{The normalized differential distribution for
  $\mathcal{R}(m_t^{pole},\rho_s)$ and $M_{be^+}$ for $\sqrt{s}=13$ TeV with
	$\mu_0 = m_t$ for five different values of the top quark mass.}
  \label{fig:mtdep}
\end{figure}

Let us now discuss the extraction of the top quark mass parameter from the
shapes of the above mentioned differential distributions. To illustrate their
sensitivity to the $m_t$ parameter we show in Fig.~\ref{fig:mtdep} the two
observables for five different values of the top quark mass. Thus, fitting the
shape of the distribution as a function of the top quark mass to a measured
distribution allows one to extract the top quark mass parameter.
We redirect the reader to Ref.~\cite{Bevilacqua:2017ipv} for a description
of the statistical analysis.
We employ the full off-shell calculation using a top quark mass of
$m_t^{in} = 173.2$ GeV, the scale $\mu_0 = H_T/2$ and the CT14 PDF set in order to
generate pseudo-data sets for an assumed integrated luminosity of
$\mathcal{L}=2.5~\rm{fb}^{-1}$ and $\mathcal{L}=25~\rm{fb}^{-1}$.
Then, template distributions for five different top quark masses between
$168.2$ GeV and $178.2$ GeV for different scales and PDF sets are generated
for the three different approaches discussed earlier: \textit{Full}, \textit{NWA}
and \textit{NWA$_{prod}$}. These templates will be used to fit the shape of the
pseudo-data set as a function of $m_t$. The minimum of the $\chi^2$ distribution
of the fit yields the extracted top quark mass $m_t^{out}$. In
order to avoid statistical fluctuations we repeat this procedure $1000$ times.

\begin{table}
  \includegraphics[width=\textwidth]{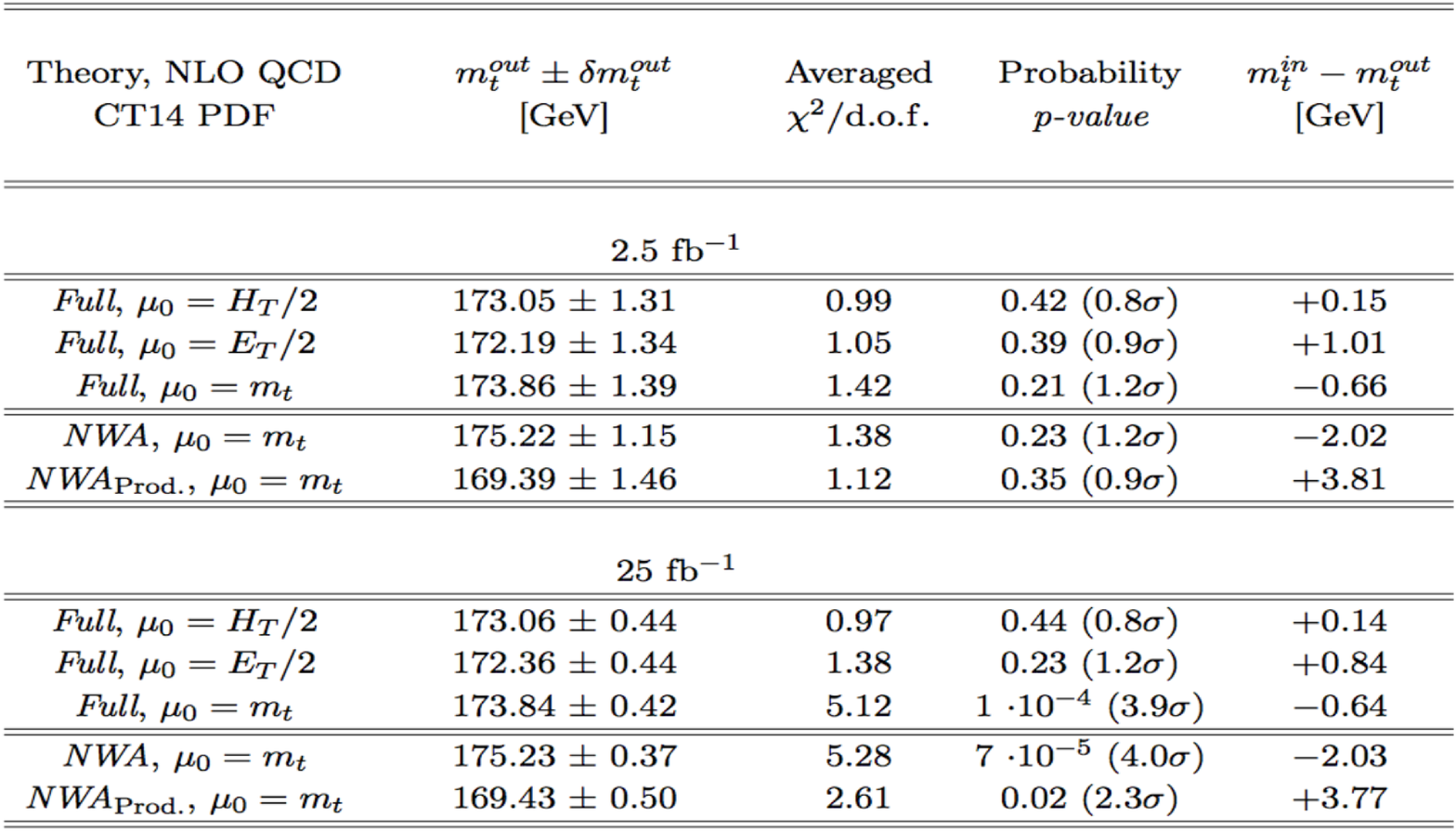}
  \caption{Results of the template fits for the $\mathcal{R}(m_t^{pole},\rho_s)$
  distribution for the LHC at $\sqrt{s}=13$ TeV. $m_t^{out}$ and
  $\delta m_t^{out}$ refer to the mean and the $1\sigma$ statistical uncertainty of
  the $1000$ pseudo-data experiments. The averaged $\chi^2/dof$ and the p-value
  indicate the quality of the performed fit. In the last column, the mass shift
  $m_t^{in}-m_t^{out}$ with $m_t^{in}=173.2$ GeV is presented.}
  \label{tab:fit_rho}
\end{table}
In Tab.~\ref{tab:fit_rho} the results of the top quark mass extraction from
the $\mathcal{R}(m_t^{pole},\rho_s)$ distribution for
$\mathcal{L}=2.5~\rm{fb}^{-1}$ and $\mathcal{L}=25~\rm{fb}^{-1}$ are shown.
Let us first note that the fit using templates from the full off-shell
calculation with the $H_T/2$ scale reproduces exactly the input top quark mass
of $173.2$ GeV, which yields a good cross check that the fitting procedure works.
Using the full off-shell calculation but different central scales $\mu_0$ for
the templates yields a shift in the extracted top quark mass of the order of
$1$ GeV, as can be seen in the last column of Tab.~\ref{tab:fit_rho}.
Furthermore, if we compare the obtained mass shifts for \textit{Full}
and \textit{NWA} for the fixed scale $\mu_0 = m_t$ we see,
independent of the considered integrated luminosity, a difference of $1.4$ GeV,
which has to be attributed to off-shell effects and continuum contributions.
Adressing scale uncertainties by a simultaneous rescaling of the
renormalization and factorization scale yields an uncertainty of $0.6-1.2$ GeV
for the dynamical scales and $2.1-2.8$ GeV for the fixed scale.
The uncertainty related to different PDF sets amounts to $0.4-0.7$ GeV.
\begin{table}
  \includegraphics[width=\textwidth]{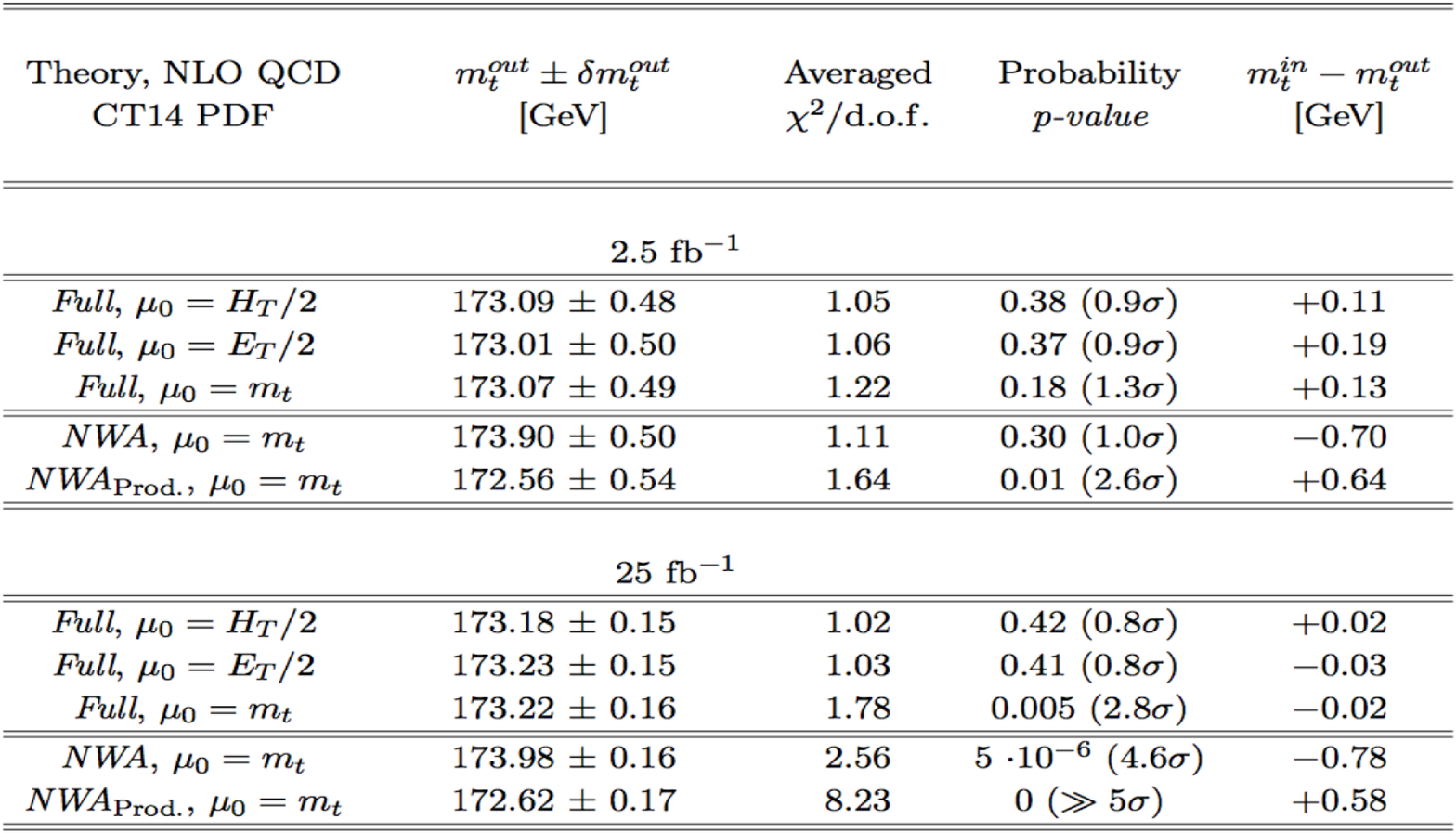}
  \caption{Results of the template fits for the $M_{be^+}$
  distribution for the LHC at $\sqrt{s}=13$ TeV. $m_t^{out}$ and
  $\delta m_t^{out}$ refer to the mean and the $1\sigma$ statistical uncertainty of
  the $1000$ pseudo-data experiments. The averaged $\chi^2/dof$ and the p-value
  indicate the quality of the performed fit. In the last column, the mass shift
  $m_t^{in}-m_t^{out}$ with $m_t^{in}=173.2$ GeV is presented.}
  \label{tab:fit_Mbl}
\end{table}
In Tab.~\ref{tab:fit_Mbl}, the corresponding results obtained for the normalized
$M_{be^+}$ distribution are shown. Overall, one can say that this observable
is more sensitive to the top quark mass parameter than the
$\mathcal{R}(m_t^{pole},\rho_s)$ distribution
by looking at the resulting statistical uncertainty on the extracted top quark
mass $\delta m_t^{out}$, the fit quality as well as the obtained mass shifts.
The choice of the central scale $\mu_0$ does not have a strong impact on the
extracted top quark mass and only amounts to a $10$ MeV uncertainty, as can be
seen from the mass shifts of the full off-shell calculation. The impact of
off-shell and non-resonant contributions on the extracted top quark mass amounts
to a mass shift of around $800$ MeV which is perfectly consistent with the
results obtained in the recent publication~\cite{Heinrich:2017bqp}.
Uncertainties related to missing higher order corrections are estimated
to be of the order of $50$ MeV for dynamical scales and $1$ GeV for the fixed
scale. PDF uncertainties are also very small, namely of the order of $30$ MeV.
\section{Conclusion}
We have studied the impact of off-shell effects on the top quark mass
extraction using template distributions. To this end we performed a systematic
comparison at fixed-order NLO QCD between the full off-shell $t\bar{t}j$
calculation and
the description of the process within the NWA. In this contribution, we focused
on two observables to highlight the findings of our more detailed study in Ref.
\cite{Bevilacqua:2017ipv}, where also additional distributions are discussed.
Our findings are that off-shell effects can have an impact on the top quark
mass extraction but this question has to be answered on the case-by-case basis. 
For example,
large off-shell effects as in the case of $M_{be^+}$ do not play an important
role because they are only visible in kinematic regions which are less important
for the top quark mass extraction. On the other hand, we found off-shell effects
of less than $15\%$ in the top mass sensistive region of the
$\mathcal{R}(m_t^{pole},\rho_s)$ distribution. Therefore, these effects have an
impact on the extracted value of the top quark mass. In this case, we find that
fits based on the narrow width approximation lead to large mass shifts. 


\end{document}